\documentclass{llncs}

\usepackage{epsfig}
\begin{document}

\newcommand{\avk}{\langle k \rangle}
\newcommand{\fluck}{\langle k^2 \rangle}

\title{Traffic-driven model of the 
World Wide Web graph}

\author{Alain Barrat \inst{1}, Marc Barth{\'e}lemy \inst{2}, 
Alessandro Vespignani \inst{1}} 
\institute{Laboratoire de Physique
Th{\'e}orique (UMR du CNRS 8627), B\^atiment 210, Universit{\'e} de
Paris-Sud 91405 Orsay, France
\and 
CEA-Centre d'Etudes de
Bruy{\`e}res-le-Ch{\^a}tel, D\'epartement de Physique Th\'eorique et
Appliqu\'ee BP12, 91680 Bruy{\`e}res-Le-Ch{\^a}tel, France}

\date{\today}
\maketitle 

\begin{abstract}
We propose a model for the World Wide Web graph 
that couples the topological growth with the traffic's
dynamical evolution. The model is based on a simple traffic-driven
dynamics and generates weighted directed graphs exhibiting the statistical 
properties observed in the Web. In particular, the model yields a 
non-trivial time evolution of vertices and heavy-tail distributions 
for the topological and traffic properties. The generated
graphs exhibit a complex architecture with 
a hierarchy of cohesiveness levels similar to those observed in the
analysis of real data.
\end{abstract}

\section{Introduction}

The World Wide Web (WWW) has evolved into an immense and intricate
structure whose understanding represents a major scientific and
technological challenge.  A fundamental step in this direction is
taken with the experimental studies of the WWW graph structure in
which vertices and directed edges are identified with web-pages and
hyperlinks, respectively. These studies are based on crawlers that
explore the WWW connectivity by following the links on each discovered
page, thus reconstructing the topological properties of the
representative graph. In particular, data gathered in large scale
crawls~\cite{Barabasi:1999,Broder:2000,Kumar:2000,Adamic:2001,Laura:2003}
have uncovered the presence of a complex architecture underlying the
structure of the WWW graph.  A first observation is the {\em
small-world} property~\cite{watts98} which means that the average
distance between two vertices (measured by the length of the shortest
path) is very small. Another important result is that the WWW exhibits
a power-law relationship between the frequency of vertices and their
degree, defined as the number of directed edges linking
each vertex to its neighbors. This last feature is the signature of a
very complex and heterogeneous topology with statistical fluctuations
extending over many length scales~\cite{Barabasi:1999}.

These complex topological properties are not exclusive to the WWW and
are encountered in a wide range of networked structures belonging to
very different domains such as ecology, biology, social and
technological systems~\cite{Barabasi:2000,Amaral:2000,mdbook,psvbook}.
The need for general principles explaining the emergence of complex
topological features in very diverse systems has led to a wide array
of models aimed at capturing various properties of real
networks~\cite{Barabasi:2000,mdbook,psvbook}, including the WWW.
Models do however generally consider only the topological structure
and do not take into account the interaction strength \---the weight
of the link\--- that characterizes real networks
\cite{Granovetter,newmancoll,vicsek,Guimera:2003,Barrat:2003,Quince:2004}. 
Interestingly, recent studies of various types of weighted
networks~\cite{Barrat:2003,Garla:2003} have shown additional complex
properties such as broad distributions and non-trivial correlations of
weights that do not find an explanation just in terms of the
underlying topological structure. In the case of the WWW, it has also
been recognized that the complexity of the network encompasses not
only its topology but also the dynamics of information. Examples of
this complexity are navigation patterns, community structures,
congestions, and other social phenomena resulting from the users'
behavior~\cite{Huberman:1997,Huberman:1998}. In addition, Adamic and
Huberman~\cite{Adamic:2001} pointed out that the number of users of a
web-site is broadly distributed, showing the relevance and
heterogeneity of the traffic carried by the WWW.

In this work we propose a simple model for the WWW graph that takes
into account the traffic (number of visitors) on the hyper-links and
considers the dynamical basic evolution of the system as being driven
by the traffic properties of web-pages and hyperlinks. The model also
mimics the natural evolution and reinforcements of interactions in the
Web by allowing the dynamical evolution of weights during the system
growth. The model displays power-law behavior for the different
quantities, with non-trivial exponents whose values depend on the
model's parameters and which are close to the measured ones.  
Strikingly, the model recovers a heavy-tailed out-traffic 
distribution whatever the out-degree distribution. 
Finally we find non-trivial clustering
properties signaling the presence of hierarchy and correlations in the
graph architecture, in agreement with what is observed in real data of
the WWW.

\subsection{Related works: Existing models for the web}

It has been realized early that the traditional random graph model,
i.e. the Erd\"os-Renyi paradigm, fails to reproduce the
topological features found in the WebGraph such as the broad degree
probability distribution, and to provide a model for a dynamical
growing network. An important step in the modeling of evolving
networks was taken by Barab\'asi et
al. \cite{Barabasi:1999,Barabasi:2000b} who proposed the ingredient of
preferential attachment: at each time-step, a new vertex is
introduced and connects randomly to already present vertices with a
probability proportional to their degree.  The combined ingredients of
growth and preferential attachment naturally lead to power-law
distributed degree. Numerous variations of this model have been
formulated~\cite{psvbook} to include different features such as
re-wiring~\cite{Tadic:2001,Krapivsky:2001}, additional edges,
directionality~\cite{Dorogovtsev:2000,Cooper:2001},
fitness~\cite{Bianconi:2001} or limited information~\cite{Mossa:2002}.

A very interesting class of models that considers the main features of
the WWW growth has been introduced by Kumar et al.~\cite{Kumar:2000}
in order to produce a mechanism which does not assume the knowledge of
the degree of the existing vertices.  Each newly introduced vertex $n$
selects at random an already existing vertex $p$; for each
out-neighbour $j$ of $p$, $n$ connects to $j$ with a certain
probability $\alpha$; with probability $1-\alpha$ it connects instead
to another randomly chosen node. This model describes the growth
process of the WWW as a copy mechanism in which newly arriving
web-pages tends to reproduce the hyperlinks of similar web-pages;
i.e. the first to which they connect.  Interestingly, this model
effectively recovers a preferential attachment mechanism without
explicitely introducing it.

Other proposals in the WWW modeling include the use of the rank values
computed by the PageRank algorithm used in search engines, combined
with the preferential attachment ingredient~\cite{Pandurangan:2002},
or multilayer models grouping web-pages in different
regions~\cite{Laura:2002} in order to obtain bipartite cliques in the
network. Finally, recent models include the textual content
affinity~\cite{Menczer:2002} as the main ingredient of the WWW
evolution.

\section{Weighted model of the WWW graph}

\subsection{The WWW graph}

The WWW network can be mathematically represented as a directed graph
${\cal G}=(V,E)$ where $V$ is the set of nodes which are the web-pages
and where $E$ is the set of ordered edges $(i,j)$ which are the {\em
directed} hyperlinks ($i,j= 1,...,N$ where $N=|V|$ is the size of the
network). Each node $i \in V$ has thus an ensemble ${\cal V}_{in}(i)$
of pages pointing to $i$ (in-neighbours) and another set ${\cal
V}_{out}(i)$ of pages directly accessible from $i$
(out-neighbours). The degree $k(i)$ of a node is divided into
in-degree $k^{in}(i)= |{\cal V}_{in}(i)|$ and out-degree $k^{out}(i)=
|{\cal V}_{out}(i)|$: $k(i)=k^{in}(i)+k^{out}(i)$. The WWW has also
dynamical features in that ${\cal G}$ is growing in time, with a
continuous creation of new nodes and links.  Empirical evidence shows
that the distribution of the in-degrees of vertices follows a
power-law behavior. Namely, the probability distribution 
that a node $i$ has in-degree
$k^{in}$  behaves as  $P(k^{in})\sim
\left(k^{in}\right)^{-\gamma_{in}^k}$, with 
$\gamma_{in}^k=2.1\pm 0.1$ as indicated by
the largest data 
sample~\cite{Barabasi:1999,Broder:2000,Adamic:2001,Laura:2003}.  
The out-degrees ($k^{out}$) distribution of web-pages is also broad but 
with an exponential
cut-off, as recent data suggest~\cite{Broder:2000,Laura:2003}. 
While the in-degree
represents the sum of all hyper-links coming from the whole WWW and
can be in principle as large as the WWW itself, the out-degree is
determined by the number of hyper-links present in a single web-page 
and is thus constrained by obvious physical elements.

\subsection{Weights and Strengths}

The number of users of any given web-site is also distributed
according to a heavy-tail distribution~\cite{Adamic:2001}. This fact
demonstrates the relevance of considering that every hyper-link has a
specific weight that represents the number of users which are using
it. The WebGraph ${\cal G}(V,E)$ is thus a directed, weighted graph
where the directed edges have assigned variables $w_{ij}$ which specify
the weight on the edge connecting vertex $i$ to vertex $j$ ($w_{ij}=0$
if there is no edge pointing from $i$ to $j$). The standard
topological characterization of directed networks is obtained by the
analysis of the probability distribution $P(k^{in})$ [$P(k^{out})$]
that a vertex has in-degree $k^{in}$ [out-degree
$k^{out}$]. Similarly, a first characterization of weights is obtained
by the distribution $P(w)$ that any given edge has weight
$w$. Along with the degree of a node, a very significative measure of
the network properties in terms of the actual weights is obtained by
looking at the vertex incoming and outgoing strength defined
as~\cite{Yook:2001,Barrat:2003}
\begin{equation}
s_{i}^{out}=\sum_{j\in{\cal V}_{out}(i)}w_{ij}\ , \ 
s_{i}^{in}=\sum_{j\in{\cal V}_{in}(i)}w_{ji} \ ,
\end{equation}
and the corresponding distributions $P(s^{in})$ and $P(s^{out})$.
The strengths $s_{i}^{in}$ and $s_{i}^{out}$ of a node integrate the
information about its connectivity and the importance of the weights
of its links, and can be considered as the natural generalization of
the degree. For the Web the incoming strength represents the actual
total traffic arriving at web-page $i$ and is an obvious measure of
the popularity and importance of each web-page. The incoming strength
obviously increases with the vertex in-degree $k_i^{in}$ and usually
displays the power-law behavior $s\sim k^{\beta}$, with the exponent
$\beta$ depending on the specific network~\cite{Barrat:2003}.


\subsection{The model}

Our goal is to define a model of a growing graph that explicitly takes
into account the actual popularity of web-pages as measured by the
number of users visiting them. Starting from an initial seed of $N_0$
pages, a new node (web-page) $n$ is introduced in the system at each
time-step and generates $m$ outgoing hyper-links. In this study, we
take $m$ fixed so that the out-degree distribution is a delta
function. This choice is motivated by the empirical observation that
the distribution of the number of outgoing links is
bounded~\cite{Laura:2003} and we have checked that the results do not
depend on the precise form of this distribution as long as
$P(k^{out}(i)=k)$ decays faster than any power-law as $k$ grows.

The new node $n$ is attached to a node $i$ with probability
\begin{equation}
Prob(n \to i)=\frac{s_{i}^{in}}{\sum_{j} s_{j}^{in}}
\label{sdrive}
\end{equation}
and the new link $n \to i$ has a weight $w_{ni}\equiv w_0$. This
choice relaxes the usual degree preferential attachment and focuses on
the popularity---or strength---driven attachment in which new
web-pages will connect more likely to web-pages handling larger
traffic. This appears to be a plausible mechanism in the WWW and in
many other technological networks. For instance, in the Internet new
routers connect to other routers with large bandwidth and traffic
handling capabilities. In the airport network, new connections
(airlines) are generally established with airports having a large
passenger traffic~\cite{Barrat:2003,Barrat:2004a,Barrat:2004b}. The
new vertex is assumed to have its own initial incoming strength
$s_n^{in}=w_0$ in order to give the vertex an initial non-vanishing
probability to be chosen by vertices arriving at later time steps. 

\begin{figure}[t]
\begin{center}
\epsfig{file=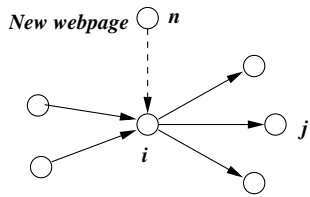,width=4cm}
\end{center}
\caption{ Illustration of the construction rule. A new web-page $n$
enters the Web and direct a hyper-link to a node $i$ with probability
proportional to $ s_{i}^{in}/\sum_j s_{j}^{in}$. The weight of the new
hyper-link is $w_0$ and the existing traffic on outgoing links of $i$
are modified by a total amount equal to $\delta_i$: $s_{i}^{out}\to
s_{i}^{out}+\delta_i$. }
\label{fig:rule}
\end{figure}

The second and determining ingredient of the model consists in considering
that a new connection $(n\to i)$ will introduce variations of the
traffic across the network. For the sake of simplicity we limit
ourselves to the case where the introduction of a new incoming link on
node $i$ will trigger only local rearrangements of weights on the
existing links $(i\to j)$ where $j\in{\cal V}_{out}(i)$ as
\begin{equation}
w_{ij}\to w_{ij}+\Delta w_{ij},
\end{equation}
where $\Delta w_{ij}$ is a function of $w_{ij}$ and of the
connectivities and strengths of $i$.  In the following we focus on the
case where the addition of a new edge with weight $w_0$ induces a
total increase $\delta_i$ of the total outgoing traffic and where this
perturbation is proportionally distributed among the edges according
to their weights [see Fig.~(\ref{fig:rule})]
\begin{equation}
\Delta w_{ij}=\delta_i \frac{w_{ij}}{s_{i}^{out}} \ .
\end{equation}
This process reflects the fact that new visitors of a web-page will
usually use its hyper-links and thus increase its outgoing
traffic. This in turn will increase the popularity of the web-pages
pointed by the hyperlinks. In this way the popularity of each page
increases not only because of direct link pointing to it but also due
to the increased popularity of its in-neighbors.  It is possible to
consider heterogeneous $\delta_i$ distributions depending on the local
dynamics and rearrangements specific to each vertex, but for the sake
of simplicity we consider the model with $\delta_i=\delta$. We finally
note that the quantity $w_0$ sets the scale of the weights. We can
therefore use the rescaled quantities $w_{ij}/w_0$, $s_i/w_0$ and
$\delta/w_0$, or equivalently set $w_0=1$. The model then depends only
on the dimensionless parameter $\delta$. The generalization to
arbitrary $w_0$ is simply obtained by replacing $\delta$, $w_{ij}$,
$s_i^{out}$ and $s_i^{in}$ respectively by $\delta/w_0$, $w_{ij}/w_0$,
$s_i^{out}/w_0$ and $s_i^{in}/w_0$ in all results.

\subsection{Analytical solution}

Starting from an initial seed of $N_0$ nodes, the network grows with
the addition of one node per unit time, until it reaches its final
size $N$. In the model, every node has exactly $m$ outgoing links with
the same weight $w_0=1$.  During the growth process this symmetry is
conserved and at all times we have $s_{i}^{out}= m w_{ij} $. Indeed,
each new incoming link generates a traffic reinforcement $\Delta
w_{ij}=\delta/m$, so that $w_{ij}=w_0+k_{i}^{in}\delta /m$ is independent
from $j$ and
\begin{equation}
s_{i}^{out}=m+\delta k_i^{in} \ .
\label{sout}
\end{equation}
The time evolution of the {\em average} of $s_i^{in}(t)$ and
$k_i^{in}(t)$ of the $i$-th vertex at time $t$ can be obtained by
neglecting fluctuations and by relying on the continuous approximation
that treats connectivities, strengths, and time $t$ as continuous
variables~\cite{Barabasi:2000,mdbook,psvbook}. The dynamical
evolution of the in-strength of a node $i$ is given by the evolution
equation
\begin{equation}
\frac{ds_{i}^{in}}{dt}=m\frac{s_{i}^{in}}{\sum_{l} s_{l}^{in}}
+\sum_{j\in{\cal V}_{in}(i)}m\frac{s_{j}^{in}}{\sum_{l}
s_{l}^{in}}\delta\frac{1}{m} \ ,
\label{eq_sin_1}
\end{equation}
with initial condition $s_{i}^{in}(t=i)=1$.
This equation states that the incoming strength of a vertex $i$ 
can only increase if a new hyper-link connects directly to $i$ (first term) 
or to a neighbor vertex $j\in{\cal V}_{in}(i)$, thus inducing a 
reinforcement $\delta/m$ on the existing in-link (second term). 
Both terms are weighted by the probability that the new vertex
establishes a hyperlink with the corresponding existing vertex.
Analogously, we can write the evolution equation for  
the in-degree $k_{i}^{in}$ that  evolves only if the new link
connects directly to $i$:
\begin{equation}
\frac{dk_{i}^{in}}{dt} =m\frac{s_{i}^{in}}{\sum_l s_{l}^{in}} \ .
\label{eq_kin_1}
\end{equation}
Finally, the out-degree is constant ($k_{i}^{out}=m$) by construction.

The above equations can be written more explicitly by noting that the
addition of each new vertex and its $m$ out-links, increase the total
in-strength of the graph by the constant quantities $1+m+m\delta$
yielding at large times $\sum_{l=1}^t
s_{l}^{in}=m\left(1+\frac{1}{m}+\delta \right)t$. By inserting this
relation in the evolution equations (\ref{eq_sin_1}) and
(\ref{eq_kin_1}) we obtain
\begin{equation}
\frac{ds_{i}^{in}}{dt}=
\frac{1}{\delta+1+\frac{1}{m}}
\left( \frac{s_{i}^{in}}{t}
+\frac{\delta}{mt}
\sum_{j\in{\cal V}_{in}(i)}s_{j}^{in} \right)
\ \  \mbox{and} \ \ 
\frac{dk_{i}^{in}}{dt}=
\frac{1}{\delta+1+\frac{1}{m}} \frac{s_{i}^{in}}{t} \ .
\label{eq_s_tot}
\end{equation}
These equations cannot be explicitly solved because of the term
$\sum_{j\in{\cal V}_{in}(i)}s_{j}^{in}$ which introduces a coupling of
the in-strength of different vertices. The structure of the equations
and previous studies of similar undirected
models~\cite{Barrat:2004a,Barrat:2004b} suggest to consider the {\em
Ansatz} $s_{i}^{in}= A k_{i}^{in}$ in order to obtain an explicit
solution. Using Eq.(\ref{sout}), and $w_{ji}=s_j^{out}/m$, we can
write
\begin{equation}
s_{i}^{in}= \sum_{j\in{\cal V}_{in}(i)} w_{ji}=
k_{i}^{in}+\sum_{j\in{\cal V}_{in}(i)} \frac{\delta}{m} k_{j}^{in},
\label{sin2}
\end{equation}
and the Ansatz $s_{i}^{in}= A k_{i}^{in}$ yields
\begin{equation}
 \sum_{j\in{\cal V}_{in}(i)} s_j^{in} = \frac{m}{\delta}(A-1) s_i^{in} \ .
\end{equation}
This allows to have a closed equation for $s_{i}^{in}$ whose solution
is
\begin{equation}
s_{i}^{in}(t) = \left(\frac{t}{i}\right)^\theta \ , \ 
\mbox{with} \ \  \theta=\frac{A}{\delta+1+1/m} 
\label{eq_mf}
\end{equation}
and $k_{i}^{in}(t)=s_{i}^{in}(t)/A$, satisfying the proposed Ansatz. 
The fact that vertices are added
at a constant rate implies that the probability distribution of
$s_{i}^{in}$ is given by \cite{psvbook,Barrat:2004a,Barrat:2004b}
\begin{equation}
 P(s^{in}, t) = \frac{1}{t+N_0} \int_0^{t} \delta(s^{in} - s_{i}^{in}(t)) di, 
\label{eqdistr}
\end{equation} 
where $\delta(x)$ is the Dirac delta function. By solving the above
integral and considering the infinite size limit $t\to\infty$ we
obtain
\begin{equation}
P(s^{in}(i)=s)\sim s^{-\gamma_{in}^s} \ , \
\mbox{with} \ 
\gamma_{in}^s=1+\frac{1}{\theta}
\end{equation}
The quantities $s^{in}_i$, $k^{in}_i$ and $s^{out}_i$ are thus here
proportional, so that their probability distributions are given by
power-laws with the same exponent
$\gamma_{in}^s=\gamma_{out}^s=\gamma_{in}^k$. The explicit value of
the exponents depends on $\theta$ which itself is a function of the
proportionality constant $A$. In order to find an explicit value of
$A$ we use the approximation that on average the total in-weight will
be proportional to the number of in-links times the average weight in
the graph $<w>=\frac{1}{t m}\sum_{l}s_{l}^{out}=(\delta+1)$.  At this
level of approximation, the exponent $\theta$ varies between $m/(m+1)$
and $1$ and the power-law exponent thus varies between $2$ ($\delta
\to \infty$) and $2+1/m$ ($\delta=0$). This result points out that the
model predicts an exponent $\gamma_{in}^k\simeq 2 $ for reasonable
values of the out-degree, in agreement with the empirical findings.

\section{Numerical simulations}

Along with the previous analytical discussion we have performed numerical
simulations of the presented  graph model in order to investigate its
topological properties with a direct statistical analysis. 

\subsection{Degree and strength distributions}

\begin{figure}[thb]
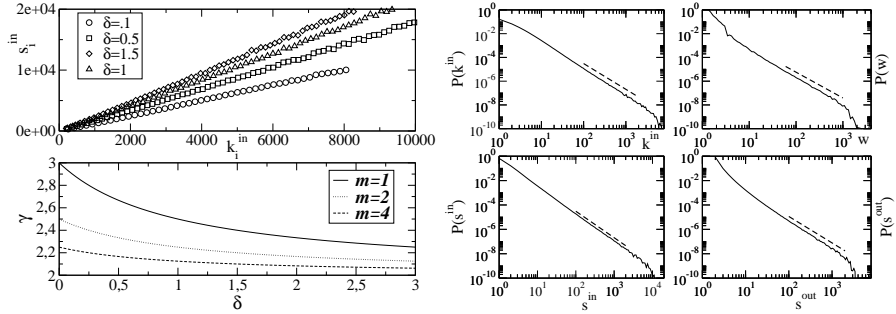

\vskip .5cm
\begin{center}
\epsfig{file=fig_svsk_gamma.eps,width=5.7cm}
\epsfig{file=fig_m2delta.5.eps,width=6cm}
\end{center}
\caption{ Top left: illustration
of the proportionality between $s^{in}$ and $k^{in}$ for various
values of $\delta$. Bottom left: theoretical approximate estimate of
the exponent $\gamma_{in}^s=\gamma_{out}^s=\gamma_{in}^k$ vs. $\delta$
for various values of $m$. Right: Probability distributions of
$k^{in}$, $w$, $s^{in}$, $s^{out}$ for $\delta=0.5$, $m=2$ and
$N=10^5$. The dashed lines correspond to a power law with exponent
$\gamma=2.17$ obtained by measuring first the slope $A$ of $s^{in}$
vs. $k^{in}$ and then using equations ($11$) and ($13$) to compute
$\gamma$. }
\label{fig:distributions}
\end{figure}

As a first test of the analytical framework we confirm numerically
that $s^{in}$, $k^{in}$, $s^{out}$ are indeed proportional and grow as
power-laws of time during the construction of the network [see
Fig.(\ref{fig:distributions})].  The measure of the proportionality
factor $A$ between $s^{in}$ and $k^{in}$ allows to compute the
exponents $\theta$ and $\gamma$, which are satisfactorily close to the
observed results and to the theoretical predictions obtained with the
approximation $A\approx <w>$.  Figure (\ref{fig:distributions}) shows
the probability distributions of the relevant quantities ($k^{in}$,
$w$, $s^{in}$, $s^{out}$) for $\delta=0.5$. All these quantities are
broadly distributed according to power-laws with the same exponent.  It
is also important to stress that the out-traffic is broadly
distributed even if the out-degree is not.

\subsection{Clustering and hierarchies}

Along with the vertices hierarchy imposed by the strength
distributions the WWW displays also a non-trivial architecture which
reflects the existence of well defined groups or communities and of
other administrative and social factors.  In order to uncover these
structures a first characterization can be done at the level of the
undirected graph representation. In this graph, the degree of a node
is the sum of its in- and out-degree ($k_i=k^{in}_i+k^{out}_i$) and
the total strength is the sum of its in- and out-strength
($s_i=s^{in}_i+s^{out}_i$). A very useful quantity is then the
clustering coefficient $c_i$ that measures the local group
cohesiveness and is defined for any vertex $i$ as the fraction of
connected neighbors couples of $i$ \cite{watts98}. The average
clustering coefficient $C=N^{-1}\sum_i c_i$ thus expresses the
statistical level of cohesiveness by measuring the global density of
interconnected vertex triplets in the network.  Further information
can be gathered by inspecting the average clustering coefficient
$C(k)$ restricted to classes of vertices with degree
$k$~\cite{vazquez02,Ravasz02}
\begin{equation}
C(k) = \frac{1}{N_k} \sum_{i/k_i=k} c_i \ ,
\end{equation}
where $N_k$ is the number of vertices with degree $k$.
In real WWW data, it has been observed that the $k$ spectrum 
of the clustering coefficient has a  highly non-trivial
behavior with a power-law decay as a function of $k$, signaling a
hierarchy in which low degree vertices belong generally to well
interconnected communities (high clustering coefficient) while hubs
connect many vertices that are not directly connected (small
clustering coefficient) \cite{eckmann,Ravasz02}.
\begin{figure}[thb]
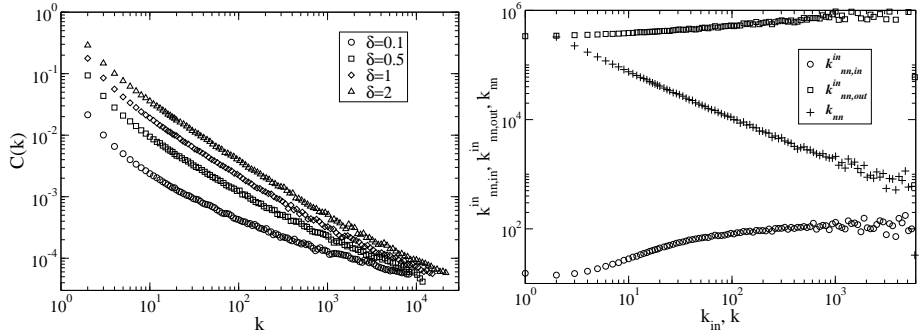

\vskip .5cm
\begin{center}
\epsfig{file=fig_ck.eps,width=6cm}
\epsfig{file=fig_knn.eps,width=6cm}
\end{center}
\caption{Left: Clustering coefficient $C(k)$, for various values
of the parameter $\delta$. Here $m=2$ and $N=10^5$. The clustering increases
with $\delta$. Right: Correlations between degrees of neighbouring
vertices as measured by $k_{nn}(k)$ (crosses), 
$k_{nn,in}^{in}(k^{in})$ (circles) and $k_{nn,out}^{in}(k^{in})$
(squares); $m=2$, $\delta=0.5$ and $N=10^5$.
}
\label{fig:ck}
\end{figure}

We show in figure (\ref{fig:ck}) the clustering coefficient $C(k)$ for
the model we propose, for various values of $\delta$. We obtain a
decreasing function of the degree $k$, in agreement with real data
observation. In addition, the range of variations spans several orders
of magnitude indicating a continuum hierarchy of cohesiveness levels
as in the analysis of Ref.~\cite{Ravasz02}.

Another important source of information about the network 
structural organization lies in the correlations of
the connectivities of neighboring
vertices~\cite{Pastor-Satorras:2001}. Correlations can be probed by 
inspecting the average degree of nearest neighbor
of a vertex $i$ 
\begin{equation}
k_{nn,i}=\frac{1}{k_i}\sum_{j \in {\cal V}(i)} k_j \ ,
\end{equation}
where the sum runs on the nearest neighbors vertices of each vertex
$i$. From this quantity a convenient measure 
to investigate the behavior of the degree
correlation function is obtained by the  average degree of the 
nearest neighbors, $k_{nn}(k)$, for vertices of degree $k$ \cite{vazquez02} 
\begin{equation}
k_{nn}(k) = \frac{1}{N_k} \sum_{i/k_i=k} k_{nn,i}.
\end{equation}
 This last quantity is related to 
the correlations between the degree of connected vertices since 
on the average it can  be  expressed as 
\begin{equation}
k_{nn}(k) = \sum_{k'} k' P(k'|k) \ .
\end{equation}
If degrees of neighboring vertices are uncorrelated, $P(k'|k)$ is only
a function of $k'$ and thus $k_{nn}(k)$ is a constant. When
correlations are present, two main classes of possible correlations
have been identified: {\em Assortative} behavior if $k_{nn}(k)$
increases with $k$, which indicates that large degree vertices are
preferentially connected with other large degree vertices, and {\em
disassortative} if $k_{nn}(k)$ decreases with $k$~\cite{Newman:2002}.

In the case of the WWW, however, the study of
additional correlation function is naturally introduced by the directed
nature of the graph. We focus on the most significative, 
the in-degree of vertices that in our model is a measure of 
their popularity ($s^{in}\sim k^{in}$). As for the undirected 
correlation, we can study the average in-degree of in-neighbours :
\begin{equation}
k_{nn,in}^{in}(i) = \frac{1}{k^{in}(i)} 
\sum_{j \in {\cal V}_{in}(i)} k^{in}(j) \ .
\end{equation}
This quantity  measures the average in-degree of the 
in-neighbours of $i$, i.e. if
the pages pointing to a given page $i$ are popular on their turn.
Moreover, relevant information comes also from
\begin{equation}
k_{nn,out}^{in}(i) = \frac{1}{k^{out}(i)} 
\sum_{j \in {\cal V}_{out}(i)} k^{in}(j) \ ,
\end{equation}
which measures the average in-degree of the out-neighbours of $i$, i.e.
the popularity of the pages to which page $i$ is pointing.
Finally, in both  cases it is possible to look at the average of this
quantity for group of vertices with in-degree $k_i^{in}$ in order 
to study the eventual assortative or disassortative behavior. 

In Figure \ref{fig:ck} we report the spectrum of $k_{nn}(k)$,
$k_{nn,in}^{in}(k^{in})$ and $k_{nn,out}^{in}(k^{in})$ in graphs
generated with the present weighted model. The undirected correlations
display a strong disassortative behaviour with $k_{nn}$ decreasing as
a power-law. This is a common feature of most technological networks
which present a hierarchical structure in which small vertices connect
to hubs. The model defined here exhibits spontaneously the
hierarchical construction that is observed in real technological
networks and the WWW. In contrast, both $k_{nn,in}^{in}(k^{in})$ and
$k_{nn,out}^{in}(k^{in})$ show a rather flat behavior signaling an
absence of strong correlations.  This indicates a lack of correlations
in the popularity, as measured by the in-degree.  The absence of
correlations in the behaviour of $k_{nn,out}^{in}(k^{in})$ is a realistic
feature since in the real WWW, vertices tend to point to
popular vertices independently of their in-degree. We also note that
$k_{nn,out}^{in}(k^{in})\gg k_{nn,in}^{in}(k^{in})$, a signature of
the fact that the average in-degree of pointed vertices is much higher
than the average in-degree of pointing vertices. This result also is a
reasonable feature of the real WWW since the average popularity of
webpages to which any vertex is pointing is on average larger than
the popularity of pointing webpages that include also the
non-popular ones.

Finally, we would like to stress that in our model the degree
correlations are to a certain extent a measure of popularity
correlations and more refined measurements will be provided by the
correlations among the actual popularity as measured by the
in-strength of vertices. We defer the detailed analysis of these
properties to a future publication, but at this stage, it is clear
that an empirical analysis of the hyperlinks traffic is strongly
needed in order to discuss in detail the WWW architecture.

\section{Conclusion}

We have presented a model for the WWW that considers the interplay
between the topology and the traffic dynamical evolution when new
web-pages and hyper-links are created. This simple mechanism produces
a non trivial complex and scale-free behavior depending on the
physical parameter $\delta$ that controls the local microscopic
dynamics. We believe that the present model might provide a general
starting point for the realistic modeling of the Web by taking into
account the coupling of its two main complex features, its topology
and its traffic.

\paragraph{Acknowledgments}
A.B and A. V. are partially funded by the 
European Commission - Fet Open project COSIN IST-2001-33555
and contract 001907 (DELIS).


\end{document}